# Metal/organic/metal bistable memory devices

*D. Tondelier, K. Lmimouni & D. Vuillaume[1]*

**Institut d'Electronique, de Microelectronique et de Nanotechnologie – CNRS
BP69, avenue Poincaré, F-59652, Villeneuve d'Ascq, France.**

*C. Fery[1] & G. Haas*

**Thomson
BP19, 1 avenue Belle Fontaine, F-35511 Cesson-Sévigné, France.**

## Abstract

We report a bistable organic memory made of a single organic layer embedded between two electrodes, we compare to the organic/metal nanoparticle/organic tri-layers device [L.P. Ma, J. Liu, and Y. Yang, *Appl. Phys. Lett. 80*, 2997 (2002)]. We demonstrate that the two devices exhibit similar temperature-dependent behaviors, a thermally-activated behavior in their low conductive state (off-state) and a slightly "metallic" behavior in their high conductive state (on-state). This feature emphasizes a similar origin for the memory effect. Owing to their similar behavior, the one layer memory is advantageous in terms of fabrication cost and simplicity.

---

[1] Authors to whom the correspondance should be addressed; electronic mail: dominique.vuillaume@iemn.univ-lille1.fr , christophe.fery@thomson.net

Organic field effect transistors, organic light emitting diodes, organic photovoltaic cells have attracted considerable attention during these last years due to their envisioned applications in low-cost, flexible, large area and lightweight organic electronics.[1] However, another important device, the organic memory, has received scarce attention, despite its important role in the electronic circuitry. The basic feature for a memory is to display a bistable behavior having two states associated with strongly different electrical resistances at the same applied voltage. Bistable effect in organic materials is known for more than 30 years[2] but it is only recently that high performance and reliable organic memory devices have been demonstrated. The Yang's group has demonstrated an organic bistable device (OBD) made with a three layers (3L) organic/metal nanoparticle/organic structure embedded between two electrodes (3L-OBD).[3-5] Bozano and coworkers have also demonstrated bistable effects in organic layers containing granular metal.[6,7] In this letter, we report the fabrication and electrical measurements of a bistable organic memory made of a single organic layer (1L) embedded between two electrodes (hereafter called 1L-OBD). We demonstrate that the two devices exhibit similar temperature-dependent behaviors, a thermally-activated behavior in their low conductive state (off-state) and a slightly "metallic" behavior (slight increase in the current when decreasing the temperature) in their high conductive state (on-state). Moreover, both our 1L-OBD and 3L-OBD exhibit a large on/off current ratio of ~$10^9$.

We fabricated several series of the two types of devices. The first device consists of a tri-layers (3L) organic/metal/organic structures embedded between two metal electrodes, following the work by the Yang's group.[3-5] The second one is simply made of a single organic layer (1L) between two electrodes, as in a MIM-like (metal-insulator-metal) device. We used pentacene, a well known material, as the organic. Both devices have a cross-bar structure. A first row of aluminum electrodes (80 nm thick) is evaporated through a shadow mask on a 250 nm thick thermally grown silicon dioxide on a silicon wafer. The silicon wafer serves as a mechanical substrate, the leakage current through the 250 nm thick $SiO_2$ is negligible (always lower than $10^{-13}$ A within the bias range used in this work). For the 3L device, we evaporated the pentacene (both lower and upper films) at 0.1 Å·s$^{-1}$ with thicknesses in the range of 50-800 nm (the same for both). The middle aluminum layer has a thickness of 15-20 nm evaporated at 2 Å·s$^{-1}$. For the 1L device, we evaporated only one layer of pentacene (150 nm to 1 µm) at an evaporation rate between 0.07 and 2.5 Å·s$^{-1}$. To complete the both type of devices, a second row of aluminum electrodes (300 nm thick), oriented perpendicularly to the first one was again evaporated at 2 Å·s$^{-1}$ through a shadow mask. The resulting device area is 1 mm$^2$. For all the electrical measurements, the bottom electrode was grounded and we applied the bias on the top electrode.

Figure 1 shows the typical current-voltage (I-V) characteristics for two devices of about the same overall thickness. Both devices exhibit the same kind of bistable behavior. They start in the off-state (the typical resistance is $R_{off}$~$10^9$-$10^{11}$ $\Omega$), they turns into a highly conductive state, the on-state, above a certain threshold voltage (its exact value being dependent on the device structure, see below). The device resistance in the on-state is very low, typically $R_{on}$< a few $\Omega$ to a few tens $\Omega$. Then, the devices remain in this on-state (second down-scaling voltage sweep on the I-V curves) until a specific action (reset) has been taken to return the device to its off-state. Applying a 0 V bias for a short period (few µs to few ms) returns the device to its off-state. The on/off ratios of our devices are very large, ~$10^9$. Since the same bistable and memory behavior is observed, the advantage of the 1L device is its more simple fabrication at a lower cost. To understand in more details the behavior, and the underlying physics, of these two types of OBD, we carried out temperature dependence measurements. Figure 2 shows the evolution of the currents as a function of the temperature. The off-state of the 1L device is thermally activated with activation energy of about 0.47-0.72 eV (sample to sample dependent). Quite similar values (0.6-0.83 eV) are found for our 3L device. The on-state shows a slightly "metallic" behavior (slight increase in the current when decreasing the temperature, inset Fig. 2). These features confirm that the behavior displayed by the original 3L device can be obtained with the more simples 1L device. Moreover, the same physical mechanisms are probably in action in both cases. The bistable effect is ascribed to be due to the inclusion of metal nanoparticules (NPs) in the organic matrix. In the 3L devices, the middle thin layer of Al provides these NPs. In the 1L device, the NPs are included in the pentacene by diffusion during the top electrode evaporation. The proof is that we have not observed a bistable effect in our 1L device if the pentacene is directly and gently contacted by a mechanical contact wire (Fig. 1). Another supporting evidence of the role of the metal NPs is that the quality of the pentacene film has no influence on the bistable behavior of the 1L device. Figure 3-a shows the I-V behaviors of 1L devices made with two different deposition rates (2.5 Å·s$^{-1}$ and 7x10$^{-2}$ Å·s$^{-1}$) for the pentacene, the rest of the device structures being identical. TM-AFM and X-ray diffractions experiments (Figs. 3-b) confirm the difference in the pentacene film morphologies. The pentacene film deposited at the lowest rate shows larger grain size (size in the range of 220-660 nm at 7x10$^{-2}$ Å·s$^{-1}$ and 50-180 nm at 2.5 Å·s$^{-1}$) in agreement with previous reports.[8] In addition, the X-ray diffraction study shows that the pentacene film deposited at the highest rate exhibits two polymorphs, the so-called "bulk" phase and "thin-film" phase, while the film deposited slowly has only the "thin-film" phase (or a bulk phase not detectable).[9] Focusing on the 1L device, we examined the memory behavior as a function of the pentacene film thickness (Fig. 4). The switching threshold voltage $V_T$ varies approximately linearly with the film thickness, it corresponds to a "switching" field of about 65-80 kV·cm$^{-1}$. Only devices with a pentacene thickness in the range 150-600 nm show the bistable behavior. Below 150 nm, the 1L-devices were always in the on-state, above 600 nm, they were always in the off-state and never switched to the on-state even at large electric field.

The bistable effect in the 1L device is attributed to the inclusion of Al NPs in the pentacene during the evaporation of the top electrode. It is well established that metal atoms can migrate inside the organic layer during the metallization where they can aggregate to form metal NPs. For instance, it was shown that such an interdiffusion phenomenon increases when decreasing the deposition rate and that the density of metal NPs increases too.[10] Thus we consider that our 1L devices consist of Al NPs included and distributed in the pentacene matrix.

The thermally activated charge transport in the off-state may be either due to a Poole-Frenkel (PF) transport between NPs or to the Schottky contact at the electrode/pentacene interface. In both cases, the theoretical energy barrier height for holes is ~ 0.7 eV (considering an ionization potential of 5 eV for pentacene and a work function of 4.3 eV for Al). In the experimental situation, this value may be reduced (force image for instance…). The experimental activation energy of 0.4-0.8 eV (Fig. 2) is consistent with these mechanisms. To distinguish between bulk-limited (PF) and interface-limited (Schottky) transports, experiments with a different nature of the NPs and the external electrodes are in progress. When increasing the applied voltage above $V_T$, the sharp increase in the current may be tentatively ascribed to the apparition of a Fowler-Nordheim (FN) tunneling between adjacent NPs as already mentioned for a similar bistable effect observed in polymer films containing metal NPs.[11] This model can be discarded here. First, the FN behavior is not in agreement with the ohmic behavior of the I-V curves in the on-state and, second, a transition from a PF to a FN regime cannot explain a dynamic range as large as $10^9$ between the off-state and on-state currents at such a low switching field (65-80 kV·cm$^{-1}$). We believe that the switching to the on-state is due to field-induced percolation of the NPs, thus forming nano-filamentary metallic pathways through the organic film. This latter mechanism is in agreement with the slight "metallic" temperature behavior observed on several devices. The relative variation of the device resistance in the on-state between 100 and 300 K, $\Delta R(100,300K)/R(300K)$, is about 0.3. For comparison, the variation of the resistivity of bulk aluminum $\Delta\rho(100,300K)/\rho(300K)$ is ~ 6. The device is not frankly shorted, this temperature variation is consistent with the presence of more or less metallic filamentary pathways made of granular metal in the pentacene film. Above $V_T$, the electric field can modify the distribution of the NPs in the organic film. This mechanism is similar to a recent report using Cu electrodes where Cu distribution in the organic field is dynamically controlled by the applied bias.[12] When we switch off the applied field, the metal NPs can relax and the device switches back to the off-state with a given time constant (reset dynamic has not been studied in details here, further works are in progress). We have obtained excellent retention time in the on-state (preliminary measurements), the "on" current (measured at 0.5 V) does not decreased after a week. More than 100 write/erase cycles have been applied without a detrimental degradation of the memory performance.

In summary, we showed that 1-layer and 3-layer organic bistable devices exhibit similar current-voltage characteristics and similar current-temperature dependences. We observed on-state current over off-state current ratios as large as $10^9$. This behavior is attributed to the inclusion of metal nanoparticules into the organic material during the top electrode evaporation for both types of devices. This behavior was observed for film thickness in the range 150-600 nm with the highest on/off ratio for the thicker films. Owing to their similar behavior, the 1-layer organic memory device is advantageous in terms of fabrication cost and simplicity.

This research has been supported by a CIFRE-grant (D.T.) from the Association Nationale de la Recherche Technique (# 448/2001). We thank R. Desfeux and A. da Costa for the X-ray diffraction measurements.

# FIGURE CAPTIONS

Figure 1: Typical current-voltage (I-V) curves for the 3L-OBD (■) with the structure pentacene (170 nm)/aluminum (20 nm)/pentacene (170 nm); for the 1L-OBD (▲) with a pentacene film of 400 nm. (●) I-V curve for a 1L-OBD (400 nm of pentacene) without the top Al electrode and mechanically contacted with a thin gold wire (the off-current is lower because the contact area in that case is smaller than with the evaporated Al contact). No switching has been observed for applied bias up to 10 V.

Figure 2: Typical Arrhenius plots of the off-state current (measured at 0.1 V) for a 3L-OBD (■) and a 1L-OBD (●) – same devices as in Fig. 1. The activation energies (straight lines are the fits) are 0.83 eV and 0.72 eV, respectively. Inset: current-temperature behaviors of the on-state current (measured at 0.1 V) for the 3L-OBD (■) and a 1L-OBD (●).

Figure 3: (a): Current-voltage behaviors for the 1L-OBD with 380-400 nm thick pentacene film evaporated at (■) 0.07 Å·s$^{-1}$ and (▲) 2.5 Å·s$^{-1}$. (b) X-ray diffraction pattern of the pentacene film at two evaporation rates. The "thin-film" phase shows 1$^{st}$ and 2$^{nd}$ order diffraction peaks at 5.74-5.76° and 11.46-11.48°, the corresponding peaks of the "bulk" phase are at 6.12° and 12.28°. Inset: Tapping mode AFM images (10 μm x 10μm) of the pentacene film sublimed at 0.07 Å·s$^{-1}$ (left image) and at 2.5 Å·s$^{-1}$ (right image).

Figure 4: Current-voltage curves of several 1L-OBD with different pentacene film thickness.

# REFERENCES

[1] J. M. Shaw and P. F. Seidler, IBM J. Res. & Dev. **45,** 3-9 (2001).

[2] J. Kevorkian, M. M. Labes, D. C. Larson, and D. C. Wu, Discuss. Faraday Soc. **N51,** 139 (1971).

[3] L. Ma, J. Liu, S. Pyo, and Y. Yang, Appl. Phys. Lett. **80,** 362-364 (2002).

[4] L. Ma, J. Liu, and Y. Yang, Appl. Phys. Lett. **80,** 2997-2999 (2002).

[5] L. Ma, S. Pyo, J. Ouyang, Q. Xu, and Y. Yang, Appl. Phys. Lett. **82,** 1419-1421 (2003).

[6] L. D. Bozano, B. W. Kean, V. R. Deline, J. R. Salem, and J. C. Scott, Appl. Phys. Lett. **84,** 607-609 (2004).

[7] M. Beinhoff, L. D. Bozano, J. C. Scott, and K. R. Carter, Polym. Mat.: Sci. Eng. **90,** 211-212 (2004).

[8] S. Pratontep and M. Brinkmann, Phys. Rev. B **69,** 165201 (2004).

[9] D. Knipp, R. A. Street, A. Völkel, and J. Ho, J. Appl. Phys. **93,** 347-355 (2003).

[10] A. C. Dürr, F. Schreiber, M. Kelsch, H. D. Carstanjen, H. Dosch, and O. H. Seeck, J. Appl. Phys. **93,** 5201-5209 (2003).

[11] C. Laurent, E. Kay, and N. Souag, J. Appl. Phys. **64,** 336-343 (1988).

[12] L. Ma, Q. Xu, and Y. Yang, Appl. Phys. Lett. **84,** 4908-4910 (2004).

Figure 1

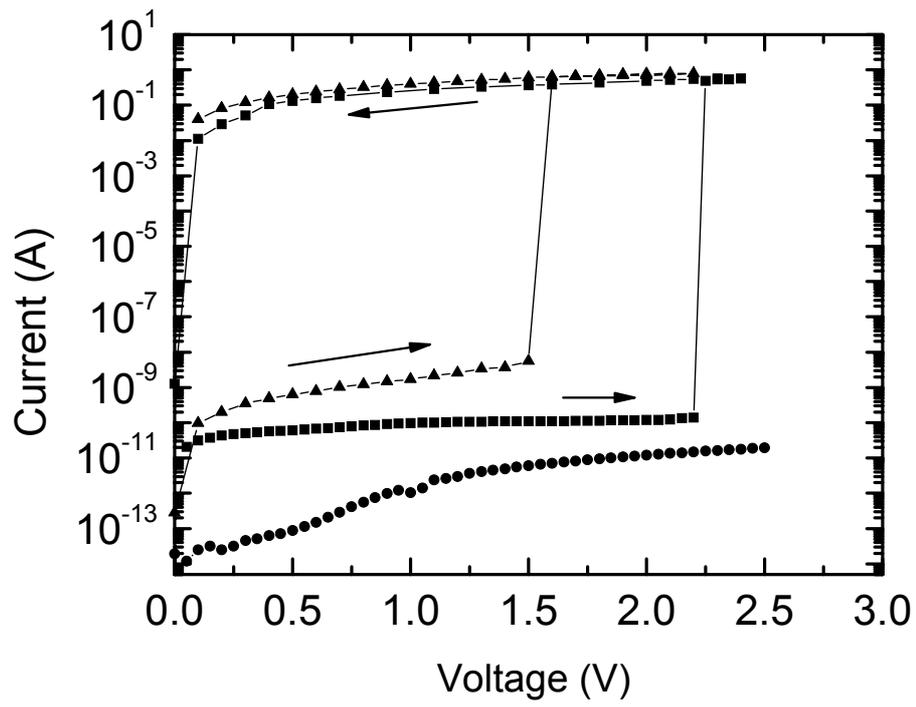

Figure 2

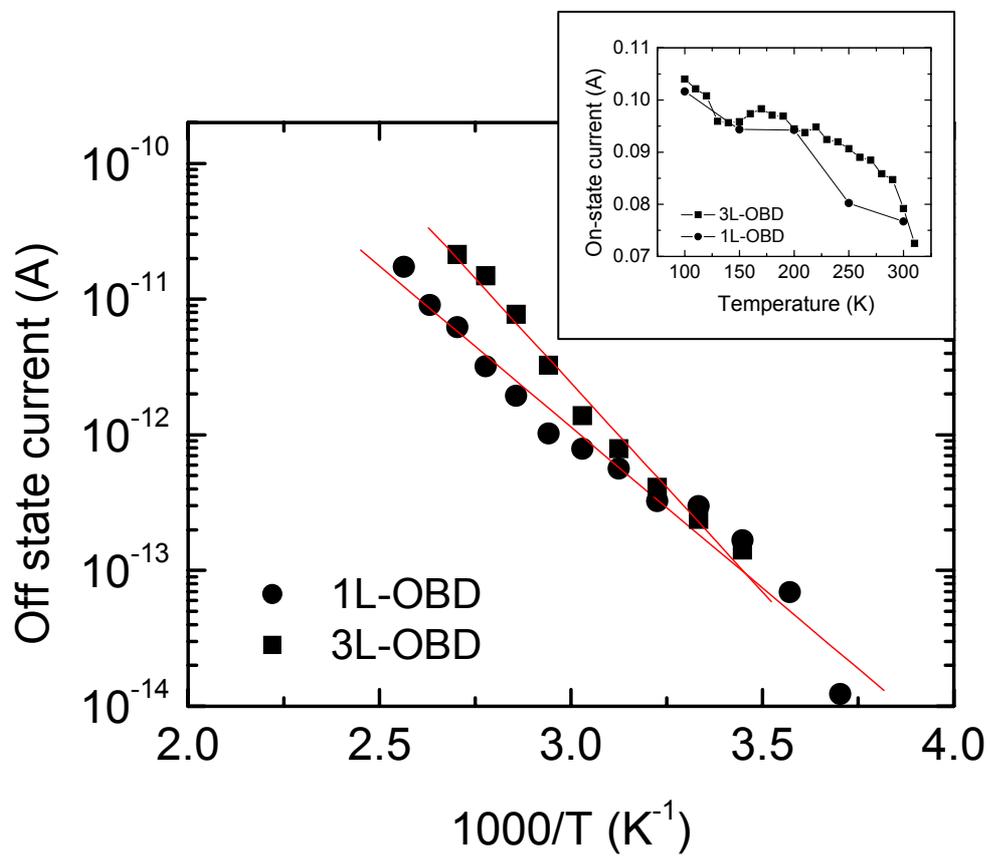

Figure 3

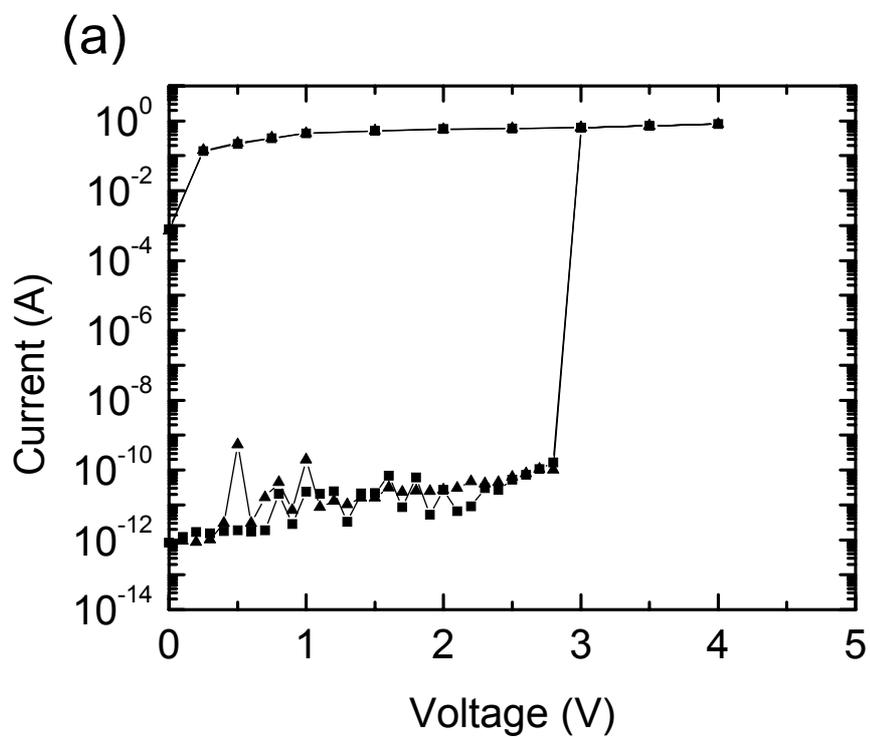

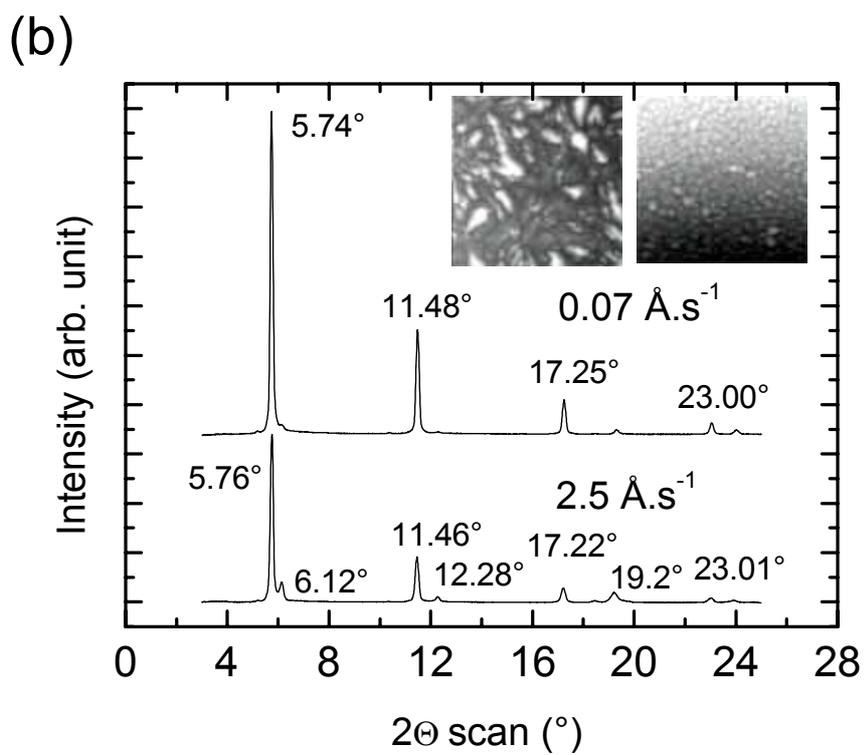

Figure 4

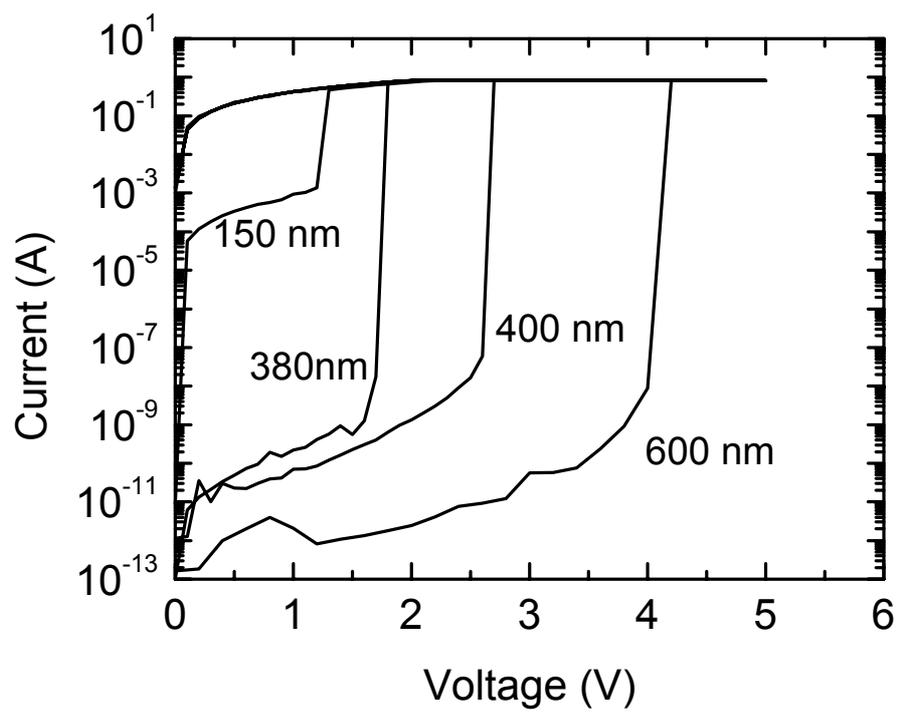